# A Comparative Analysis of Inconel 718 Made by Additive Manufacturing and Suction Casting: Microstructure Evolution in Homogenization

Yunhao Zhao, Kun Li, Matthew Gargani, Wei Xiong*

Physical Metallurgy and Materials Design Laboratory,
Department of Mechanical Engineering and Materials Science,
University of Pittsburgh, Pittsburgh, PA 15261, USA
* Corresponding Author, Email: weixiong@pitt.edu, w-xiong@outlook.com
Tel. +1 (412) 383-8092, Fax: +1 (412) 624-4846

## Abstract

Homogenization is one of the critical stages in the post-heat treatment of additive manufacturing (AM) component to achieve uniform microstructure. During homogenization, grain coarsening could be an issue to reserve strength, which requires careful design of both time and temperature. Therefore, a proper design of homogenization becomes particularly important for AM design, for which work hardening is usually no longer an option. In this work, we discovered an intriguing phenomenon during homogenization of suction-cast and AM Inconel 718 superalloys. Through both short and long-term isothermal heat treatments at 1180°C, we observed an abnormal grain growth in the suction-cast alloy but continuous recrystallization in the alloy made by laser powder bed fusion (LPBF). The grain size of AM samples keeps as small as 130 μm and is even slightly reduced after homogenization for 12 h. The homogeneity of Nb in the AM alloys is identified as the critical factor for NbC formation, which further influences the recrystallization kinetics at 1180°C. Multi-type dislocation behaviors are studied to elucidate the grain refinement observed in homogenized alloys after LPBF. This work provides a new pathway on microstructure engineering of AM alloys for improved mechanical performance superior to traditionally manufactured ones.

## 1. Introduction

As one of the most widely used additive manufacturing (AM) techniques, laser powder bed fusion (LPBF) is capable of building components with complex geometry layer-by-layer [1–6] and thus requires less capital investment in tools and dies that are essential in the conventional subtractive manufacturing [7]. LPBF has been applied to manufacture different types of engineering alloys, such as Inconel 718, which is a precipitation-hardenable superalloy with excellent mechanical performance [8–16]. However, due to extremely high heating/cooling rates and cycles, LPBF is easy to introduce anisotropic grain structure in the as-built alloys [7,8,17–22], which becomes a challenge for AM technique development. Moreover, because of the segregation of Nb in the interdendritic area, the microstructure of as-solidified Inconel 718 often contains the hexagonal

Laves_C14 phase, (Ni, Fe, Cr)$_2$(Nb, Ti, Mo) [16,23–28], which accelerates crack initiation and propagation [29], and thus reduces tensile and stress rupture properties [30]. Additionally, the Laves_C14 phase formation consumes the alloying elements such as Nb and Ti, which are essential to the formation of the γ" (Ni$_3$Nb, bct_D0$_{22}$) and γ' (Ni$_3$(Ti, Al), fcc_L1$_2$) strengthening particles. Therefore, the Laves_C14 phase needs to be dissolved by homogenization, which becomes critical in the post-heat treatment design to achieve the desired mechanical properties of the Inconel 718 [23,31–41]. Figure 1 summarizes some typical heat treatments of wrought Inconel 718 [42–44], and there are three homogenization choices: (i) 0.5~1 h at around 970°C; (ii) 2 h at around 1000°C; and (iii) 2 h at around 1065°C. In fact, a considerable amount of work has found that the above heat treatment methods are insufficient for LPBF fabricated alloys due to the anisotropic microstructures developed uniquely during laser melting [20,28,36,40]. In the LPBF fabricated alloys, columnar grain texutre usually develops due to the thermal gradient induced during the laser melting process [21,45–47]. Recently, some research studies [28,36,48–51] indicate that homogenization at a higher temperature above 1100°C could result in the better mechanical performance of Inconel 718, although the phase transformation mechanism behind this is yet unclear. Therefore, it is essential to gain insights into the microstructure evolution during homogenization for microstructure engineering on AM alloys.

As a consequence, this work aims to study phase transformation behaviors during the homogenization of LPFB-made Inconel 718. It is noteworthy that alloys under the LPFB process experience multiple cyclic heating and cooling with different heating/cooling rates depending on the laser scanning pattern, while regular casting typically has one single melting and cooling for sample preparation. Suction-casting can be regarded as an extreme case of the casting methods, as it can produce solidification rates at the order of $10^2$~$10^4$ K/s [52,53], which are much faster than those of the traditional casting methods (0.1~0.01 K/s) [54]. It can also refine the microstructure and extend the solid solubility [55]. However, the LPBF-made alloys have a much faster solidification rate at a magnitude of $10^6$ K/s [56] than the suction-cast alloys. Therefore, it is interesting to compare the microstructure evolution between the extreme casting case, i.e., the suction-casting and the LPBF method. Hence, suction-cast alloys are used as a reference for comparison in this work. Computational thermodynamics is utilized to guide the experimental design through the CALPHAD (CALculation of PHAse Diagrams) modeling.

## 2. Modeling and experiments

### 2.1. Thermodynamic modeling

Both equilibrium and nonequilibrium step diagrams (phase fraction vs. temperature) of Inconel 718 are plotted in Figs. 2(a) & (b) using the Thermo-Calc software according to the thermodynamic database TCNI8, which is used for nickel alloy study. Solid phases such as γ (fcc_A1), MC carbides (NbC with the fcc_B1 structure), δ (Ni$_3$Nb with the D0$_a$ structure), γ" and γ' are often observed during heat treatment. As shown in Fig. 2(a), the NbC carbide is stable at high temperatures, even at 1200°C, and thus often co-exists with the γ matrix phase during homogenization. With the proper size and distribution range, the NbC carbide can act as the grain boundary pinning particles to impede grain growth [22,57,58], whereas the coarsened NbC particles can introduce micro-void softening effects with localized stress concentration and crack initiation to reduce grain boundaries strength [59–61]. The δ phase is stable up to 1034°C (Fig. 2(a)). It usually forms at grain boundaries with a needle shape, owns incoherent interfaces with the γ matrix, and thus is detrimental to the alloy strength. Both γ" and γ' particles are the strengthening phases of Inconel 718. They distribute dispersively in the matrix as nano-size particles with coherent or semi-coherent interfaces with γ matrix [12,62,63]. The γ" phase is metastable and will transform into the δ phase after long-time aging [64–66]. Its phase fraction is predicted by the nonequilibrium step diagram, as shown in Fig. 2(b). The phase transformation during solidification is predicted by the nonequilibrium Scheil-Gulliver modeling [67,68]. As shown in Fig. 2(c), the NbC and Laves_C14 phases can form during solidification, although the latter one is metastable. It is noteworthy that, although the prediction indicates the late formation of δ and σ phases close to the incipient melting point during solidification, these two phases are not often observed in the as-solidified microstructure.

Although an isothermal heat treatment above the δ solvus temperature for homogenization at 1065°C shown in Fig. 1, may remove the Laves_C14 phase and avoid the δ phase formation, a higher homogenization temperature of 1180°C is considered in this work to explore new pathways for microstructure engineering. The homogenization at 1180°C can efficiently dissolve the Laves_C14 phase and promote the homogenization process. This temperature of 1180°C is above the reported recrystallization temperature of 1100°C in AM Inconel 718 alloys [48] and can thus introduce recrystallization effectively for grain refinement. However, in industrial production, it needs further evaluation to adopt 1180°C as the homogenization temperature because such a higher temperature is above the predicted incipient melting point (1110°C), as shown in Fig. 2(c). Hence, a lower homogenization temperature may be more suitable in production to avoid liquidation cracking. It should be noted that considering the model simplification with infinite diffusion in liquid and no diffusion in solid phases, the predicted incipient point of 1110°C by Scheil-Gulliver modeling is lower than the experimental value of 1165°C, which was reported for the as-cast samples [69]. However, such a difference is acceptable by considering the approximation applied in the Scheil-Gulliver model [67,68].

## 2.2. Experiments

One rod shape sample of Inconel 718 with a diameter of 11 mm and a length of 40 mm was prepared by suction-casting under a pure argon atmosphere using an arc-melter (ABJ-338, manufactured by Materials Research Furnaces Inc.). AM Inconel 718 samples were built by LPBF using an EOS M 290 machine with default laser melting parameters [70] with the laser power of 285 W, the scan velocity of 960 mm/s, and the hatching space of 0.11 mm. The hatch lines between adjacent layers have a rotation angle of 67°. The compositions of the suction-cast alloy and the AM alloy are listed in Table 1, which shows the composition of the two alloys are very close. Both alloys were sectioned into parts and encapsulated into vacuumed quartz tubes with back-filled pure argon. Afterward, samples were homogenized at 1180°C for 20 min, 1 h, or 12 h followed by quenching in ice-water. Table 2 is a list of notations for eight samples under different homogenization conditions.

After the heat treatment, all eight samples were surface polished followed by microstructure analysis with electron microscopy. SEM (scanning electron microscopy, Zeiss Sigma 500 VP, Carl Zeiss AG) and EDS (Oxford Instruments plc) characterizations were carried out for microstructure morphology observation and composition determination. Phase fractions were estimated by analyzing the SEM images using the ImageJ software package. EBSD (electron backscatter diffraction, FEI Scios Dual-Beam, FEI Company) was used to study recrystallization process of the homogenized samples with a mapping area of 1200 μm × 1200 μm and step size of 1.2 μm.

The EBSD results were analyzed by the OIM Analysis™ v8 software package. TEM (transmission electron microscope) samples were mechanically thinned to about 50 μm, and further polished using an automatic twin-jet electropolisher (Model 110, E.A. Fischione Instruments, Inc.) with a voltage of 15 V at -30°C. The electropolishing solution is a mixture of 10 vol.% perchloric acid and 90 vol.% methanol. TEM characterization was conducted on H-9500 E-TEM (Hitachi, Ltd.) under an acceleration voltage of 300 kV.

## 3. Results and discussion

### 3.1. Microstructures of as-cast and as-built samples

Microstructure evolution in suction-cast and LPBF alloys during homogenization are shown in Figure 3 as SEM-BSE (backscattered electron) images for comparison. As shown in Fig. 3(a), in the as-cast sample, the grains formed during the solidification are equiaxial with a diameter between 10 and 20 μm. Irregular-shaped Nb-rich Laves_C14 phase forms along grain boundaries as the predominant precipitate. A small fraction of blocky NbC carbides is encompassed by the Laves_C14 precipitates, as shown in Fig. 3(a), which is consistent with the Scheil-Gulliver solidification model-prediction. As predicted by Scheil-Gulliver simulation in Fig. 3(c), the formation temperature of NbC carbides (1298°C) is higher than that of the Laves_C14 phase (1170°C). However, the δ phase can be observed neither in the alloys by suction-casting nor LPBF.

Figure 3(e) shows the as-built microstructure of the X-Z plane (along building direction Z) of the as-built sample. Due to the large directional thermal gradient introduced by laser melting, columnar grains grow along the direction normal to the edge of single arc-shaped melt pools, as depicted by yellow chained lines in Fig. 3(e). It should be noted that both very fine columnar (with a width of ~1 μm) and cellular (with a diameter of ~2 μm) subgrains can be found within the relatively larger grains of the overlapped melt pool areas, as indicated in Fig. 3(e). The only secondary phase observed in the as-built sample is the Laves_C14 phase, which forms along grain boundaries of both the columnar and cellular subgrains due to Nb segregation. The phases are further confirmed using XRD (X-ray diffraction). As discussed earlier, Nb is of vital importance for the formation of strengthening phases, while its microsegregation leads to the formation of the detrimental Laves_C14 phase. Hence, it is critical to control the Nb segregation during homogenization to ensure a high Nb homogeneity in the γ matrix to promote strengthening precipitation such as γ'' and γ' phases in subsequent heat treatment steps. The LPBF can produce alloys with a better initial Nb homogeneity due to the solute trapping effect, which is caused by the rapid solid-liquid interface velocity during fast solidification in the laser melting process [71].

### 3.2. Phase transformations in homogenized alloys

Phase transformation behaviors during homogenization at 1180°C in both suction-cast and AM alloys can be analyzed through Fig. 3. For the suction-cast alloys, 20-min and 1-hour homogenization at 1180°C can dissolve most of the Laves_C14 phase in the as-cast microstructure and remain the NbC carbides formed during solidification. However, some remaining Laves_C14 particles can still be observed near the NbC carbides, as shown in Figs. 3(b)&(c). After 12-hour homogenization (Fig. 3(d)), the Laves_C14 phase dissolves completely, and only NbC carbides can be observed as precipitates. For the AM alloys, no NbC carbide can be observed in the as-built alloy, which is probably due to its precipitation is suppressed by the fast solidification rate. 20-min homogenization (Fig. 3(f)) causes the Laves_C14 phase formed in the as-built condition to dissolve into the γ matrix completely. Meanwhile, only a very small amount of NbC carbides can be found to form along grain boundaries in sample AM20m. Microstructure evolution shown in Figs. 3(f)-3(h) demonstrates that the homogenizations at 1180°C for a long time can promote more NbC carbide formation. A notable increase of the particle size of NbC carbides can be identified in sample AM12h (Fig. 3(h)) after 12-hour homogenization by comparing with samples AM20m (20-min, Fig. 3(f)) and AM1h (1-hour, Fig. 3(g)).

Table 3 summarizes the fractions of Nb-rich phases in suction-cast and AM alloys. Since both Nb-rich phases, NbC carbide and Laves_C14 phase, form together, it is challenging to differentiate them for statistical analysis. Instead, the total fraction of the Nb-rich phases was estimated according to the SEM-BSE image to quantify the difference between suction-cast and AM alloys. The total phase fraction of the Nb-rich phases decreases during the homogenization process in suction-cast alloys, while it increases for the AM alloys. After 12-hour homogenization, the Nb-rich phase fraction of suction-cast alloys becomes close to the one of the AM alloys and approaches a constant, 0.47%, which is consistent with the CALPHAD model-prediction on the amount of equilibrium NbC showed in Fig. 2(a). This is because an extended homogenization time will cause the two alloys to approach equilibrium states. However, since there are plenty of NbC carbides in the as-cast alloy, whereas the higher solidification rate results in essentially the absence of NbC carbides in the as-built alloy, the fraction of Nb-rich phases in the suction-cast alloy is decreasing, but in AM alloys it is increasing.

From the analysis of the phase transformations above, it can be speculated how the Nb homogeneity change in both alloys during homogenization. In the as-cast sample, the Laves_C14 phase usually locates near NbC carbides. Due to the dissolution of the Laves_C14 phase, as can be observed in samples AC20m (Fig. 3(b)) and AC1h (Fig. 3(c)), the Nb will be released, and dissolve into the γ matrix, hence the Nb homogeneity around NbC particles should increase. However, for the AM samples, owing to the increase of NbC carbides from almost 0% in sample AM20m to 0.36% in sample AM12h (Table 3), it can be inferred that the formation of NbC is due to the accumulation of Nb near the NbC nuclei. Such a process will decrease the Nb homogeneity. Consequently, the homogeneity of the Nb content in the Inconel 718 alloy, which only contains 5 wt.% Nb (i.e., 3.4 at.% Nb), can be significantly influenced by the homogenization process. It should be noted that the initial homogeneity of Nb in Inconel 718 can affect its distribution in the homogenized samples, and can further influence the precipitation of Nb-rich phases (such as γ" and δ phases) in the subsequent isothermal/athermal processes. Therefore, it is essential to further evaluate the evolution of Nb homogeneity during homogenization processes for the optimization of post-heat treatment of Inconel 718.

In order to further quantify the Nb homogeneity, the Nb concentration distribution of the γ matrix in the vicinity of NbC carbides was determined by EDS point identification. For each sample, three NbC carbides were studied, and an average value was taken to represent the Nb homogeneity of the sample. There are three steps in the EDS point identification, which is illustrated in Fig. 4(a): (i) a 10 μm×10 μm square matrix is identified with the position of a typical NbC particle at the center; (ii) the EDS point identification was performed for Nb concentrations on 24 nodes which are evenly distributed within the square, the step size between each node is 2.5 μm. (iii) the average Nb concentration for each node was calculated for contour diagram plots shown in Fig. 4(b). According to this, it is observed that the general trend of the evolution of Nb homogeneity for suction-cast alloys decreases with homogenization time, whereas for AM alloys the trend is the opposite.

### *3.3. Grain size evolution: recrystallization and precipitates-pinning effects*

Figures 5-7 present the grain size evolution and the recrystallization behaviors in both suction-cast and AM alloys, which were observed under EBSD. It should be pointed out that because of the large grain size observed in samples AC1h and AC12h, two extra EBSD scans with mapping areas of 2500 μm × 2500 μm were conducted in addition to the 1200 μm × 1200 μm mapping areas used for all samples. The grain characteristics such as average grain size and grain size distribution of samples AC1h and AC12h are calculated from the 2500 μm × 2500 μm mappings. As depicted in Fig. 5, for the suction-cast alloys, the average grain size increases significantly with increased homogenization duration. However, the grain size evolution is opposite in the AM samples. Surprisingly, the grain size in the sample AM12h after 12-hour isothermal treatment at 1180°C is even smaller than AM20m, which is heat-treated for 20 min.

In the left column of Fig. 6, i.e., Figs. 6(a)-(e), the microstructure evolution of suction-cast alloys during isothermal heat treatment at 1180°C is presented based on the analysis of the inverse pole figure (IPF) orientation maps. The IPF orientation maps with 2500 μm × 2500 μm mapping areas for samples AC1h and AC12h are shown in the subfigures of Figs. 6(d)&(e), respectively. Figure 6(a) shows the evolution of grain size distribution for suction-cast alloys with the as-cast state (AC in Fig. 6(b)) and homogeneous states at 1180°C for 20 min (AC20m in Fig. 6(c)), 1 h (AC1h in Fig. 6(d)), and 12 h (AC12h in Fig. 6(e)). The area-weighted distribution is used in this work because the long-time homogenization leads to coarsened grains and reduced grain numbers in cast alloy; therefore, it can reflect the relationship between grain characteristics and materials properties in a more suitable way. The unimodal grain size distribution of sample AC20m shown in Fig. 6(a) indicates a uniform grain growth during the homogenization of the as-cast samples at 1180°C for 20 min with the microstructure evolution from Fig. 6(b) to 6(c). After 1-hour homogenization, a bimodal grain size distribution shown in Fig. 6(a) indicates a discontinuous grain growth. The left peak of the grain size distribution curve of AC1h implies the initial grains, and the right peak represents the grown grains from AC20m. This can be observed through a comparison between Figs. 6(c) and 6(d). As illustrated in Fig. 6(e), significantly coarsened grains can be found in sample AC12h with some small grains remained at the triple junctions of these coarsened grains. Correspondingly, a trimodal grain size distribution can also be observed in sample AC12h (Fig. 6(a)), of which two peaks appearing with smaller grain diameters of 150 μm and 400 μm. The distinct large and small grains observed in sample AC12h (Fig. 6(e)) indicate an abnormal grain growth during homogenization.

According to Hillert [72], the abnormal grain growth can be initiated by a continuous decrease of a factor $z = 3f/4r$, which represents the dispersion level of secondary phase particles [73]. The $f$ and $r$ represent the phase fraction and particle size of the secondary phase, respectively. The decrease of phase fraction and/or increase of particle size of the secondary phase will lead to a decrease of the dispersion level $z$ and thus reduce the pinning effects of the particles. When there are much larger grains existing in such circumstances, abnormal grain growth can occur. Accordingly, in our case, the occurrence of abnormal grain growth found in suction-cast alloys during homogenization should be attributed to the continuous dissolution of the Laves_C14 phase.

Using the method introduced by Kusama et al. [74], the abnormal grain growth rate is estimated and compared with experimental data. The details of the calculation are given in the appendix. The calculation shows the experimental abnormal grain growth rate in sample AC12h is $4.7 \times 10^{-9}\ m/s$, which is much slower than the theoretical prediction of $2.067 \times 10^{-5}\ m/s$, indicating the particles of NbC carbides and Laves_C14 phase uniformly distributed can drag the grain boundaries, and thus retard their movement.

In the middle column of Fig. 6, i.e., Figs. 6(f)-(j), the microstructure evolution of AM alloys during isothermal heat treatment at 1180°C is presented based on the analysis of the IPF orientation maps. Figure 6(f) shows the evolution of grain size distribution for AM alloys made by LPBF with the as-built state (AB in Fig. 6(g)) and homogenized state at 1180°C for 20 min (AM20m in Fig. 6(h)), 1 h (AM1h in Fig. 6(i)), and 12 h (AM12h in Fig.6(j)). In Fig. 6(f), the small peak at a grain diameter of 50 µm in the grain size distribution curve of AM20m indicates the initiation of recrystallization in sample AM20m. After 1-hour homogenization, three distinct peaks can be observed in the grain size distribution curve for sample AM1h (Fig. 6(f)). The average grain size of sample AM1h (Fig. 5) also becomes a smaller value of 128 µm. After 12-hour homogenization, the average grain size value in sample AM12h decreases further down to 113 µm (Fig. 5). The grain size distribution of AM12h is close to unimodal (Fig. 6(f)). The grain size is not significantly increased after 12-hour homogenization at 1180°C from sample AM1h to sample AM12h, especially when compared with that in the cast alloys. This is likely due to the Zener pinning effects introduced by the NbC particles [72].

The Zener pinning effect [57,58,72,75] of the fine carbide particles, NbC, at the grain boundaries is confirmed by the TEM characterization on sample AM12h, as shown in Fig. 6(k). It is found that dispersed NbC particles distribute along and near grain boundaries, indicating that the fine NbC carbides precipitating along with recrystallization at or near grain boundaries can impede grain boundary movement effectively and hence refine the grain size. Contrarily, no pinning effect was found in suction-cast alloys mainly due to the large NbC particle size in the early stage of homogenization and their average distribution inside of the matrix of the grains.

Grain orientation spread (GOS) generated from EBSD analysis is useful for quantifying microstructure attributes during recrystallization [76–86]. The average GOS [86] of one grain $i$ is defined as

$$\text{GOS}(i) = \frac{1}{J(i)} \sum_j \omega_{ij} \tag{1}$$

of which $J(i)$ represents the pixel numbers of the grain $i$, $\omega_{ij}$ is the misorientation angle between the orientation of pixel $j$ and the mean orientation of grain $i$.

The GOS value represents the degree of grain distortion. A larger distortion within one grain leads to a higher GOS value. The GOS value for recrystallized grains should be low since less stored energy saved in these grains [79]. As reviewed by Hadadzadeh et al. [79], the threshold of the GOS can be selected between 1~5° for identifying recrystallized grains. Nevertheless, the GOS values are relatively low in the present study, since the samples are free from external deformation. Therefore, it is hard to determine the recrystallized grains by solely defining a threshold of the GOS. Alternatively, the recrystallization behaviors can be studied through a combined analysis of both grain size and GOS. The GOS distribution and mapping diagrams of the homogenized samples are presented in Fig. 7. The GOS maps with 2500 µm × 2500 µm mapping areas for samples AC1h and AC12h are shown in the subfigures of Figs. 7(d)&(e), respectively. The area-weighted GOS distribution is used following the same reasons proposed in grain size distribution analysis.

A further investigation on the GOS of suction-cast alloys (Figs. 7(a)-(e)) shows that it experiences an increase from sample AC to sample AC20m, and then decreases slightly in sample AC1h. The increase of GOS in sample AC20m can be attributed to the strains introduced by grain growth and dissolution processes of the Laves_C14 phase, which reduced from 13.13% in the as-cast state to 0.73% in the 20-min homogenized state (Table 3). With 1-hour homogenization, both GOS and phase fraction of Laves_C14 get reduced. After 12-hour homogenization, the GOS in sample AC12h increases, which is due to the increased grain size [86] caused by abnormal grain growth.

The GOS evolution observed in AM alloys experiences a decrease from sample AM20m to AM1h, while it slightly increases in sample AM12h, as shown in Figs. 7(f)-(j). Sample AM20m has relatively high GOS values, probably because of the distortion accumulation due to residual stress introduced during the LPBF process [31]. Sample AM1h shows the lowest and uniform GOS distribution (Figs. 7(f)&(i)), indicating almost all grains are continuously recrystallized. The GOS in sample AM12h (Figs. 7(f)&(j)) increases slightly compared with sample AM1h, whereas it is still at a low level. The reason for such a change can be attributed to the deformation induced by Nb accumulation to form more NbC carbides with significantly increased particle size, as shown in Table 3 and Fig. 3(h).

### 3.4. Dislocation behaviors in homogenized Inconel 718

For the homogenized AM alloys, a large number of dislocations were found in the vicinity of NbC carbides and grain boundaries. The interactions of dislocation-carbide and dislocation-grain boundary can influence the precipitation and grain boundary hardening remarkably. Figure 8 shows multiple dislocation behaviors observed in sample AM12h by TEM. A considerable amount of parallel dislocation arrays/loops are present in the vicinity of larger NbC carbides around 200~300 nm in diameter. However, the dislocation arrays/loops surrounding small NbC carbides are rarely found. This may be because small NbC carbides have coherency with γ matrix, as demonstrated by the selected area electron diffraction shown in the inset of Fig. 8(b). The small NbC carbides have similar diffraction patterns to the matrix, with the preferred orientation relationship of $(0\bar{2}2)_{NbC}[\bar{2}33]_{NbC}//(1\bar{1}1)_{\gamma}[\bar{1}12]_{\gamma}$. The comparison of the number of dislocations between carbides with various sizes indicates the generation of dislocations around large NbC carbides results from the loss of coherency of NbC/γ phase boundaries during the growth of carbides particles, as illustrated in Figs. 8(a)&(c), which is consistent with other reported experiments [87–89]. Furthermore, the relief of residual stress during the homogenization process may generate local strains inside of the matrix, while once the local strain field encounters impediment from NbC carbides, dislocations can be emitted to reduce the energy [90]. These two factors contribute to the generation of dislocations and make large NbC carbides be the dislocation sources. Direct evidence of NbC carbides being dislocation sources is noted in Fig. 8(a) by the yellow arrow (i.e., the one on the right-hand-side pointing to the left), of which dislocation loops are emitted from NbC carbides with a larger size, about 200 nm in diameter.

The dislocations in sample AM12h mainly appear in planar arrays (Figs. 8(a),(c)&(d)), yet tanglings and cells caused by a cross slip of dislocations are seldom found. Such a situation agrees with the observation from the work of Sundararaman et al. [91], where the reason lies in the intermediate stacking fault energy of Inconel 718 (50~70 mJ/m$^2$) [91,92], which is sufficiently low to inhibit cross slip during homogenization. However, the stacking fault energy of Inconel 718 is not such low to form stacking faults bounded with separated partial dislocations [91]. Likewise, no such stacking fault structure is found in the present work. In fact, when applying external stresses, dislocation tanglings and cells can be formed by cross slip, as reported in the work of Zhang et al. [60]. The present and previous studies indicate the dislocation behaviors can be changed depending on various conditions for the range of stacking fault energy in Inconel 718.

The interactions between dislocations and precipitates in sample AM12h are found to follow both Orowan bowing and dislocation cutting mechanisms. As indicated by the red arrows (the first three counted from the left-hand-side) in Fig. 8(a), dislocations cut through small particles and bow around large particles while moving to grain boundaries. The dislocation arrays are impeded by grain boundaries then pile up (Figs. 8(a)&(d)). According to Kondo et al. [93], the interactions between dislocations and grain boundaries are related to the rotation of the Burgers vector when the dislocation is to cross the grain boundaries, leaving a residual dislocation. The formation of the residual dislocation requires additional energy, making the process energetically unfavorable. Consequently, dislocations are prone to pile up at grain boundaries. Further dislocation pileup leads to new dislocation generation in the adjacent grain, as marked by the red arrows in Fig. 8(d), the activation of new dislocations is able to release local strains and lower the possibility of propagation of cracks, which affects the hardening of the alloys. The aforementioned interactions of dislocations/NbC carbides/grain boundaries also result in the subsequent Zener pinning effect on grain boundaries, which is beneficial to the recrystallization and grain refinement.

## 4. Conclusions

- This work presents a comparative analysis of the microstructure evolution at 1180°C in Inconel 718 alloys made by suction-cast and AM-LPBF. The initial microstructures of the as-cast and as-built samples are very different, such a difference can influence the Nb homogeneity of the matrix, which is found to be a critical factor for grain morphology development during homogenization. The Nb homogeneity increases in suction-cast alloys, whereas it decreases in AM alloys with the time increases during isothermal heat treatment at 1180°C.
- Phase transformations during homogenization in suction-casting are dissimilar to the ones in AM alloys mainly due to divergent initial solidification microstructure. In suction-cast alloys, the Laves_C14 phase remains during 1-hour short-time homogenization and dissolves after 12-hour long-time homogenization. In AM alloys, the Laves_C14 phase dissolves completely within 20-min homogenization accompanied by the growth of the NbC carbide, which continues in the entire homogenization process.
- Abnormal grain growth is observed in suction-cast alloys with increased homogenization time, while the grain growth is impeded, and the average grain size is even refined in AM alloys for longer homogenization durations. The recrystallization process is found to be continuous for the AM alloys. The grain refinement is attributed to Zener pinning effect by NbC carbides.
- Multi-types of dislocation behaviors are observed in homogenized AM alloys, which may affect the hardening of the alloys. NbC carbides losing coherency with the matrix are identified as dislocation sources. The growth of NbC carbides and the release of local strains during homogenization processes are two major factors contributing to the dislocation generation.
- Dislocations generated in the homogenized AM alloys are found to be arrays/loops, which can be interpreted by the intermediate stacking fault energy of Inconel 718. Both Orowan bowing

    and dislocation cutting mechanisms are found regarding the dislocation-precipitate interaction. The pileup of dislocations at grain boundaries can activate new dislocation arrays in the adjacent grain, which is conducive to local strain relief and preventing initiation of cracks to harden the alloys.
- This work indicates a well-designed homogenization for Inconel alloys is essential during the post-heat treatment for AM processes. In addition, the homogenization above the incipient melting of the Inconel 718 can successfully reduce the grain texture inherited from the LPBF process.

## Acknowledgments

The authors are grateful for the financial support from the National Aeronautics and Space Administration under the Grant Number (NNX17AD11G) and for the support from Thermo-Calc company on the CALPHAD modeling using the software and databases. Wei Xiong thanks Prof. Em. John Ågren at KTH Royal Institute of Technology, Dr. Qing Chen at the Thermo-Calc Software AB, Sweden, and Dr. Soumya Sridar for the valuable discussion. The authors also thank Dr. Bingtao Li, Dr. Shuying Chen, Miss. Yinxuan Li, and Miss. Xiaoyi Liu for their help on sample preparation.

## Appendix A Calculation and experimental results of abnormal grain growth rate for the suction-cast sample homogenized at 1180°C for 12-h, AC12h

According to Kusama et al. [74], the grain growth rate can be expressed as:

$$\frac{dR}{dt} = M^{gb} \cdot \Delta G \qquad (1)$$

of which $R$ is the radius of a grain, $t$ is the time, $M^{gb}$ is the grain boundary mobility, $\Delta G$ is the driving pressure of the grain growth.

The $\Delta G$ includes two contributions: $\Delta G_s$, which represents the driving pressure from subgrain boundaries, and $\Delta G_h$, which is the driving pressure from pre-existing high-angle grain boundaries. In this work, no subgrains are observed. Hence $\Delta G$ becomes:

$$\Delta G = \Delta G_h = \sigma_h V_m \left(\frac{C_n}{R_n} - \frac{C_a}{R_a}\right) \qquad (2)$$

where $\sigma_h$ is the grain boundary energy, which is not readily available. The only reported value for the grain boundary energy in Inconel 718 is considered as 0.424 J/m² from Nishimoto et al. [94] during the study of grain boundary liquation. This value is comparable with other work for the Inconel alloys [95]. $V_m$ is the molar volume of Inconel 718, and is calculated to be $7.38 \times 10^{-6} \ m^3/mol$ using the TCNI8 database released by the Thermo-Calc software AB. $R_n$ and $R_a$ are the mean radii of normal grains and abnormal grains, respectively. The value of $R_n$ is taken as 113.5 $\mu m$ (representing the left peak of grain size distribution in sample AC1h, Fig. 6(i)), and $R_a$ is estimated to be 406.6 $\mu m$ (representing the weighetd average grain size of abnormal grains in sample AC12h). $C_n$ and $C_a$ are constants, and are taken as 1.5 and 1, respectively, for 3D growth case and when $R_n \ll R_a$ [74]. The value of $\Delta G$ is calculated as 0.0337 J/mol.

The grain boundary mobility $M^{gb}$ is calculated by:

$$M^{gb} = \frac{D^{gb}}{\delta RT} \qquad (3)$$

where $\delta$ is the grain boundary thickness and is used as $5 \times 10^{-10} \ m$ [74,96,97], $R$ is the constant of the ideal gas, $T$ is the temperature, i.e., 1453.15 K (1180°C). $D^{gb}$ is the diffusivity of the grain boundary. Due to the lack of experimental data of $D^{gb}$ in Inconel 718, experimental parameters measured by Cermak [96] in a Ni-10.04Fe-18.98Cr (in wt.%) alloy are adopted. The $D^{gb}$ is estimated by [96,97]:

$$D^{gb} = \frac{1}{s \cdot \delta} \{s\delta D^{gb}\}_0 \exp\left(-\frac{Q^{gb}}{RT}\right) \qquad (4)$$

where $s$ is the segregation factor, which equals to the ratio of the concentrations of an element $i$ in the grain boundary and in the matrix near the grain boundary. It is found that when the content of Cr approaches to about 20 wt.%, the gradients of elemental concentrations near grain boundaries are small, and the values of $s$ of Cr and Fe can be considered as 1 [96]. In our case, the concentration of Cr in Inconel 718 is about 19 wt.%, which is close to the threshold of 20 wt.% in [96]. Additionally, it is reasonable to speculate that the high-temperature and long-time homogenization applied on the alloys can reduce the grain boundary segregation to a relatively low level. Therefore, $s = 1$ is an acceptable assumption. $\{s\delta D^{gb}\}_0$ is the pre-exponential factor and $Q^{gb}$ is the activation energy of grain boundary self-diffusion [96,97]. The values of $\{s\delta D^{gb}\}_0$ and $Q^{gb}$ are taken as $1.349 \times 10^{-10}\ m^3/s$ and $2.187 \times 10^5\ J/mol$ from [96], respectively. These values are for the diffusion of Fe in the grain boundaries of the Ni-10.04Fe-18.98Cr (in wt.%) alloy, since the diffusion of Fe was found to be slower than that of Cr, as reported in [96], becoming a controlling factor of grain boundary movement. The $D^{gb}$ is calculated to be $3.71 \times 10^{-9}\ m^2/s$.

The $M^{gb}$ is estimated to be $6.142 \times 10^{-4}\ mol \cdot m/(J \cdot s)$. Hence, the theoretically predicted abnormal grain growth rate is $dR/dt = 2.067 \times 10^{-5}\ m/s$.

The experimental abnormal grain growth rate for the suction-cast sample homogenized at 1180°C for 12 h, i.e., the sample AC12h, is determined as follows. The grain growth from the beginning (0 h) to 12 h can be described by:

$$R(t)^2 - R_0^2 = kt \qquad (5)$$

where $R(t)$ is the average abnormal grain radius at time $t$, and $R_0$ is the initial grain radius at the time $t_0$. In this case, $R(t)$ is $R_a$, and $R_0$ is the average grain radii (16 $\mu m$) in the as-cast sample, respectively. $k$ is a constant and calculated as $3.82 \times 10^{-12}\ m^2/s$, and $t = 12\ h$. Thus the experimental value of $\frac{dR}{dt}(t = 12\ h)$ is $4.7 \times 10^{-9}\ m/s$.

## Appendix B Supplementary data

Supplementary material related to this article can be found, in the online version, at doi:https://doi.org/10.1016/j.addma.2020.101404.

# References

[1]  S. Das, Physical aspects of process control in selective laser sintering of metals, Adv. Eng. Mater. 5 (2003) 701–711. https://doi.org/10.1002/adem.200310099.

[2]  J.P. Kruth, P. Mercelis, J. Van Vaerenbergh, L. Froyen, M. Rombouts, Binding mechanisms in selective laser sintering and selective laser melting, Rapid Prototyp. J. 11 (2005) 26–36. https://doi.org/10.1108/13552540510573365.

[3]  K. Osakada, M. Shiomi, Flexible manufacturing of metallic products by selective laser melting of powder, Int. J. Mach. Tools Manuf. 46 (2006) 1188–1193. https://doi.org/10.1016/j.ijmachtools.2006.01.024.

[4]  I. Yadroitsev, A. Gusarov, I. Yadroitsava, I. Smurov, Single track formation in selective laser melting of metal powders, J. Mater. Process. Technol. 210 (2010) 1624–1631. https://doi.org/10.1016/j.jmatprotec.2010.05.010.

[5]  D.D. Gu, W. Meiners, K. Wissenbach, R. Poprawe, Laser additive manufacturing of ceramic components: materials, processes, and mechanisms, Laser Addit. Manuf. Mater. Des. Technol. Appl. 6608 (2016) 163–180. https://doi.org/10.1016/B978-0-08-100433-3.00006-3.

[6]  J. Smith, W. Xiong, W. Yan, S. Lin, P. Cheng, O.L. Kafka, G.J. Wagner, J. Cao, W.K. Liu, Linking process, structure, property, and performance for metal-based additive manufacturing: computational approaches with experimental support, Comput. Mech. 57 (2016) 583–610. https://doi.org/10.1007/s00466-015-1240-4.

[7]  Z. Wang, K. Guan, M. Gao, X. Li, X. Chen, X. Zeng, The microstructure and mechanical properties of deposited-IN718 by selective laser melting, J. Alloys Compd. 513 (2012) 518–523. https://doi.org/10.1016/j.jallcom.2011.10.107.

[8]  K.N. Amato, S.M. Gaytan, L.E. Murr, E. Martinez, P.W. Shindo, J. Hernandez, S. Collins, F. Medina, Microstructures and mechanical behavior of Inconel 718 fabricated by selective laser melting, Acta Mater. 60 (2012) 2229–2239. https://doi.org/10.1016/j.actamat.2011.12.032.

[9]  F. Theska, A. Stanojevic, B. Oberwinkler, S.P. Ringer, S. Primig, On conventional versus direct ageing of Alloy 718, Acta Mater. 156 (2018) 116–124. https://doi.org/10.1016/j.actamat.2018.06.034.

[10] N. Raghavan, R. Dehoff, S. Pannala, S. Simunovic, M. Kirka, J. Turner, N. Carlson, S.S. Babu, Numerical modeling of heat-transfer and the influence of process parameters on tailoring the grain morphology of IN718 in electron beam additive manufacturing, Acta Mater. 112 (2016) 303–314. https://doi.org/10.1016/j.actamat.2016.03.063.

[11] A. Cruzado, B. Gan, M. Jiménez, D. Barba, K. Ostolaza, A. Linaza, J.M. Molina-Aldareguia, J. Llorca, J. Segurado, Multiscale modeling of the mechanical behavior of IN718 superalloy based on micropillar compression and computational homogenization, Acta Mater. 98 (2015) 242–253. https://doi.org/10.1016/j.actamat.2015.07.006.

[12] G.P. Dinda, A.K. Dasgupta, J. Mazumder, Texture control during laser deposition of nickel-based superalloy, Scr. Mater. 67 (2012) 503–506. https://doi.org/10.1016/j.scriptamat.2012.06.014.

[13] R. Cozar, A. Pineau, Morphology of γ' and γ" precipitates and thermal stability of Inconel 718 type alloys, Metall. Trans. 4 (1973) 47–59. https://doi.org/10.1007/BF02649604.

[14] M. Sundararaman, P. Mukhopadhyay, S. Banerjee, Some aspects of the precipitation of metastable intermetallic phases in INCONEL 718, Metall. Trans. A. 23 (1992) 2015–

2028. https://doi.org/10.1007/BF02647549.

[15] C.M. Kuo, Y.T. Yang, H.Y. Bor, C.N. Wei, C.C. Tai, Aging effects on the microstructure and creep behavior of Inconel 718 superalloy, Mater. Sci. Eng. A. 510–511 (2009) 289–294. https://doi.org/10.1016/j.msea.2008.04.097.

[16] D.D. Keiser, H.L. Brown, Review of the Physical Metallurgy of Alloy 718, U.S. Department of Energy, 1976. https://doi.org/10.2172/4016087.

[17] Q. Jia, D. Gu, Selective laser melting additive manufacturing of Inconel 718 superalloy parts: densification, microstructure and properties, J. Alloys Compd. 585 (2014) 713–721. https://doi.org/10.1016/j.jallcom.2013.09.171.

[18] J. Strößner, M. Terock, U. Glatzel, Mechanical and microstructural investigation of nickel-based superalloy IN718 manufactured by selective laser melting (SLM), Adv. Eng. Mater. 17 (2015) 1099–1105. https://doi.org/10.1002/adem.201500158.

[19] M. Ma, Z. Wang, X. Zeng, Effect of energy input on microstructural evolution of direct laser fabricated IN718 alloy, Mater. Charact. 106 (2015) 420–427. https://doi.org/10.1016/j.matchar.2015.06.027.

[20] S. Raghavan, B. Zhang, P. Wang, C.N. Sun, M.L.S. Nai, T. Li, J. Wei, Effect of different heat treatments on the microstructure and mechanical properties in selective laser melted INCONEL 718 alloy, Mater. Manuf. Process. 32 (2017) 1588–1595. https://doi.org/10.1080/10426914.2016.1257805.

[21] F. Yan, W. Xiong, E.J. Faierson, Grain structure control of additively manufactured metallic materials, Mater. 10 (2017) 1260-1271. https://doi.org/10.3390/ma10111260.

[22] F. Yan, W. Xiong, E. Faierson, G.B. Olson, Characterization of nano-scale oxides in austenitic stainless steel processed by powder bed fusion, Scr. Mater. 155 (2018) 104–108. https://doi.org/10.1016/j.scriptamat.2018.06.011.

[23] X. Li, J.J. Shi, C.H. Wang, G.H. Cao, A.M. Russell, Z.J. Zhou, C.P. Li, G.F. Chen, Effect of heat treatment on microstructure evolution of Inconel 718 alloy fabricated by selective laser melting, J. Alloys Compd. 764 (2018) 639–649. https://doi.org/10.1016/j.jallcom.2018.06.112.

[24] Y. Murata, M. Morinaga, N. Yukawa, H. Ogawa, M. Kato, Solidification structures of Inconel 718 with microalloying elements, Superalloys 718. (1994) 81–88.

[25] S.G.K. Manikandan, D. Sivakumar, K. Prasad Rao, M. Kamaraj, Laves phase in alloy 718 fusion zone-microscopic and calorimetric studies, Mater. Charact. 100 (2015) 192–206. https://doi.org/10.1016/j.matchar.2014.11.035

[26] A. Strondl, R. Fischer, G. Frommeyer, A. Schneider, Investigations of MX and $\gamma'/\gamma''$ precipitates in the nickel-based superalloy 718 produced by electron beam melting, Mater. Sci. Eng. A. 480 (2008) 138–147. https://doi.org/10.1016/j.msea.2007.07.012.

[27] G.D.J. Ram, A.V. Reddy, K.P. Rao, G.M. Reddy, Microstructure and mechanical properties of Inconel 718 electron beam welds, Mater. Sci. Technol. 21 (2005) 1132–1138. https://doi.org/10.1179/174328405X62260.

[28] J. Schneider, B. Lund, M. Fullen, Effect of heat treatment variations on the mechanical properties of Inconel 718 selective laser melted specimens, Addit. Manuf. 21 (2018) 248–254. https://doi.org/10.1016/j.addma.2018.03.005.

[29] D. Zhang, Z. Feng, C. Wang, W. Wang, Z. Liu, W. Niu, Comparison of microstructures and mechanical properties of Inconel 718 alloy processed by selective laser melting and casting, Mater. Sci. Eng. A 724 (2018) 357–367. https://doi.org/10.1016/j.msea.2018.03.073.

Yunhao Zhao, Kun Li, Matthew Gargani, Wei Xiong, *Additive Manufacturing*, 36 (2020) 101404
[30] G.D.J. Ram, A.V. Reddy, K.P. Rao, G.M. Reddy, Improvement in stress rupture properties of inconel 718 gas tungsten arc welds using current pulsing, J. Mater. Sci. 40 (2005) 1497–1500. https://doi.org/10.1007/s10853-005-0590-2.

[31] B. Song, S. Dong, Q. Liu, H. Liao, C. Coddet, Vacuum heat treatment of iron parts produced by selective laser melting: microstructure, residual stress and tensile behavior, Mater. Des. 54 (2014) 727–733. https://doi.org/10.1016/j.matdes.2013.08.085.

[32] X. Huang, M.C. Chaturvedi, N.L. Richards, Effect of homogenization heat treatment on the microstructure and heat-affected zone microfissuring in welded cast alloy 718, Metall. Mater. Trans. A. 27 (1996) 785–790. https://doi.org/10.1007/BF02648966.

[33] C. Radhakrishna, K.P. Rao, The formation and control of Laves phase in superalloy 718 welds, J. Mater. Sci. 32 (1997) 1977–1984. https://doi.org/10.1023/A:1018541915113.

[34] W.J. Sames, F.A. List, S. Pannala, R.R. Dehoff, S.S. Babu, The metallurgy and processing science of metal additive manufacturing, Int. Mater. Rev. 61 (2016) 315–360. https://doi.org/10.1080/09506608.2015.1116649.

[35] M. Garibaldi, I. Ashcroft, J.N. Lemke, M. Simonelli, R. Hague, Effect of annealing on the microstructure and magnetic properties of soft magnetic Fe-Si produced via laser additive manufacturing, Scr. Mater. 142 (2018) 121–125. https://doi.org/10.1016/j.scriptamat.2017.08.042.

[36] E. Chlebus, K. Gruber, B. Kuźnicka, J. Kurzac, T. Kurzynowski, Effect of heat treatment on the microstructure and mechanical properties of Inconel 718 processed by selective laser melting, Mater. Sci. Eng. A. 639 (2015) 647–655. https://doi.org/10.1016/j.msea.2015.05.035.

[37] D. Zhang, Z. Feng, C. Wang, W. Wang, Z. Liu, W. Niu, Comparison of microstructures and mechanical properties of Inconel 718 alloy processed by selective laser melting and casting, Mater. Sci. Eng. A. 724 (2018) 357–367. https://doi.org/10.1016/j.msea.2018.03.073.

[38] T. Trosch, J. Strößner, R. Völkl, U. Glatzel, Microstructure and mechanical properties of selective laser melted Inconel 718 compared to forging and casting, Mater. Lett. 164 (2016) 428–431. https://doi.org/10.1016/j.matlet.2015.10.136.

[39] D. Deng, J. Moverare, R.L. Peng, H. Söderberg, Microstructure and anisotropic mechanical properties of EBM manufactured Inconel 718 and effects of post heat treatments, Mater. Sci. Eng. A 693 (2017) 151–163. https://doi.org/10.1016/j.msea.2017.03.085.

[40] W.M. Tucho, P. Cuvillier, A. Sjolyst-Kverneland, V. Hansen, Microstructure and hardness studies of Inconel 718 manufactured by selective laser melting before and after solution heat treatment, Mater. Sci. Eng. A 689 (2017) 220–232. https://doi.org/10.1016/j.msea.2017.02.062.

[41] V.A. Popovich, E.V. Borisov, A.A. Popovich, V.S. Sufiiarov, D.V. Masaylo, L. Alzina, Impact of heat treatment on mechanical behaviour of Inconel 718 processed with tailored microstructure by selective laser melting, Mater. Des. 131 (2017) 12–22. https://doi.org/10.1016/j.matdes.2017.05.065.

[42] High Temp Metals, https://www.hightempmetals.com/techdata/hitempInconel718data.php (accessed November 3, 2019).

[43] Special Metals, https://www.specialmetals.com/assets/smc/documents/inconel_alloy_718.pdf (accessed November 3, 2019).

Yunhao Zhao, Kun Li, Matthew Gargani, Wei Xiong, *Additive Manufacturing*, 36 (2020) 101404
[44] ASTM International, B673: Specification for Precipitation-Hardening and Cold Worked Nickel Alloy Bars, Forgings, and Forging Stock for Moderate or High Temperature Service, 2018, pp. 1–7. https://doi.org/10.1520/B0637-18.2.

[45] X. Wang, L.N. Carter, B. Pang, M.M. Attallah, M.H. Loretto, Microstructure and yield strength of SLM-fabricated CM247LC Ni-Superalloy, Acta Mater. 128 (2017) 87–95. https://doi.org/10.1016/j.actamat.2017.02.007.

[46] L. Thijs, F. Verhaeghe, T. Craeghs, J. Van Humbeeck, J.P. Kruth, A study of the microstructural evolution during selective laser melting of Ti-6Al-4V, Acta Mater. 58 (2010) 3303–3312. https://doi.org/10.1016/j.actamat.2010.02.004.

[47] K. V. Yang, Y. Shi, F. Palm, X. Wu, P. Rometsch, Columnar to equiaxed transition in Al-Mg(-Sc)-Zr alloys produced by selective laser melting, Scr. Mater. 145 (2018) 113–117. https://doi.org/10.1016/j.scriptamat.2017.10.021.

[48] W.M. Tucho, V. Hansen, Characterization of SLM-fabricated Inconel 718 after solid solution and precipitation hardening heat treatments, J. Mater. Sci. 54 (2019) 823–839. https://doi.org/10.1007/s10853-018-2851-x.

[49] V.P. Sabelkin, G.R. Cobb, T.E. Shelton, M.N. Hartsfield, D.J. Newell, R.P. O'Hara, R.A. Kemnitz, Mitigation of anisotropic fatigue in nickel alloy 718 manufactured via selective laser melting, Mater. Des. 182 (2019) 108095. https://doi.org/10.1016/j.matdes.2019.108095.

[50] X. Li, J.J. Shi, G.H. Cao, A.M. Russell, Z.J. Zhou, C.P. Li, G.F. Chen, Improved plasticity of Inconel 718 superalloy fabricated by selective laser melting through a novel heat treatment process, Mater. Des. 180 (2019). https://doi.org/10.1016/j.matdes.2019.107915.

[51] W. Huang, J. Yang, H. Yang, G. Jing, Z. Wang, X. Zeng, Heat treatment of Inconel 718 produced by selective laser melting: microstructure and mechanical properties, Mater. Sci. Eng. A 750 (2019) 98–107. https://doi.org/10.1016/j.msea.2019.02.046.

[52] L. Yu, Z. Wang, J. Wu, X. Meng, X. Lai, Microstructure and properties of vacuum cast Sc-containing Be–Al alloys, Int. J. Met. 13 (2019) 201–212. https://doi.org/10.1007/s40962-018-0249-9.

[53] Y. Liu, L. Shi, G. Yang, W.Z. Jin, J. Cui, Microstructure and metastable phase in rapidly solidified TiAl alloy prepared by vacuum suction casting, Cryst. Res. Technol. 54 (2019) 1–7. https://doi.org/10.1002/crat.201900054.

[54] L.Y. Sheng, W. Zhang, J.T. Guo, L.Z. Zhou, H.Q. Ye, Microstructure evolution and mechanical properties' improvement of NiAl-Cr(Mo)-Hf eutectic alloy during suction casting and subsequent HIP treatment, Intermetallics. 17 (2009) 1115–1119. https://doi.org/10.1016/j.intermet.2009.05.003.

[55] L. Sheng, W. Zhang, J. Guo, H. Ye, Microstructure and mechanical properties of Hf and Ho doped NiAl-Cr(Mo) near eutectic alloy prepared by suction casting, Mater. Charact. 60 (2009) 1311–1316. https://doi.org/10.1016/j.matchar.2009.06.005.

[56] Y. Tian, J.A. Muñiz-Lerma, M. Brochu, Nickel-based superalloy microstructure obtained by pulsed laser powder bed fusion, Mater. Charact. 131 (2017) 306–315. https://doi.org/10.1016/j.matchar.2017.07.024.

[57] D.M. Collins, B.D. Conduit, H.J. Stone, M.C. Hardy, G.J. Conduit, R.J. Mitchell, Grain growth behaviour during near-γ' solvus thermal exposures in a polycrystalline nickel-base superalloy, Acta Mater. 61 (2013) 3378–3391. https://doi.org/10.1016/j.actamat.2013.02.028.

[58] X. Wang, Z. Huang, B. Cai, N. Zhou, O. Magdysyuk, Y. Gao, S. Srivatsa, L. Tan, L.

Jiang, Formation mechanism of abnormally large grains in a polycrystalline nickel-based superalloy during heat treatment processing, Acta Mater. 168 (2019) 287–298. https://doi.org/10.1016/j.actamat.2019.02.012.

[59] W. Xiong, G.B. Olson, Cybermaterials: materials by design and accelerated insertion of materials, npj Comput. Mater. 2 (2016) 15009. https://doi.org/10.1038/npjcompumats.2015.9.

[60] D. Zhang, W. Niu, X. Cao, Z. Liu, Effect of standard heat treatment on the microstructure and mechanical properties of selective laser melting manufactured Inconel 718 superalloy, Mater. Sci. Eng. A 644 (2015) 32–40. https://doi.org/10.1016/j.msea.2015.06.021.

[61] L. Chang, W. Sun, Y. Cui, R. Yang, Influences of hot-isostatic-pressing temperature on microstructure, tensile properties and tensile fracture mode of Inconel 718 powder compact, Mater. Sci. Eng. A 599 (2014) 186–195. https://doi.org/10.1016/j.msea.2014.01.095.

[62] M.C. Chaturvedi, Y. Han, Strengthening mechanisms in Inconel 718 superalloy, Met. Sci. 17 (1983) 145–149. https://doi.org/10.1179/030634583790421032.

[63] L. Xiao, D.L. Chen, M.C. Chaturvedi, Shearing of γ" precipitates and formation of planar slip bands in Inconel 718 during cyclic deformation, Scr. Mater. 52 (2005) 603–607. https://doi.org/10.1016/j.scriptamat.2004.11.023.

[64] Z. Chen, R.L. Peng, J. Moverare, P. Avdovic, J.M. Zhou, S. Johansson, Surface integrity and structural stability of broached Inconel 718 at high temperatures, Metall. Mater. Trans. A Phys. Metall. Mater. Sci. 47 (2016) 3664–3676. https://doi.org/10.1007/s11661-016-3515-6.

[65] Y.T. Chen, A.C. Yeh, M.Y. Li, S.M. Kuo, Effects of processing routes on room temperature tensile strength and elongation for Inconel 718, Mater. Des. 119 (2017) 235–243. https://doi.org/10.1016/j.matdes.2017.01.069.

[66] M. Dehmas, J. Lacaze, A. Niang, B. Viguier, TEM study of high-temperature precipitation of delta phase in inconel 718 alloy, Adv. Mater. Sci. Eng. 2011 (2011). https://doi.org/10.1155/2011/940634.

[67] G.H. Gulliver, The quantitative effect of rapid cooling upon the constitution of binary alloys, J. Inst. Met. 9 (1913) 120–157.

[68] E. Scheil, Bemerkungen zur schichtkristallbildung, Z. Met. 34 (1942) 70–72.

[69] W.D. Cao, R.L. Kennedy, M.P. Willis, Differential thermal analysis (DTA) study of the homogenization process in alloy 718, Superalloys 718 (1991) 147–160. https://doi.org/10.7449/1991/Superalloys_1991_147_160.

[70] Q. Chen, X. Liang, D. Hayduke, J. Liu, L. Cheng, J. Oskin, R. Whitmore, A. To, Aninherentstrain based multiscale modeling framework for simulating part-scale re-sidual deformation for direct metal laser sintering, Addit. Manuf. 28 (2019) 406–418, https://doi.org/10.1016/j.addma.2019.05.021.

[71] Y. Tian, D. McAllister, H. Colijn, M. Mills, D. Farson, M. Nordin, S. Babu, Rationalization of microstructure heterogeneity in INCONEL 718 builds made by the direct laser additive manufacturing process, Metall. Mater. Trans. A 45 (2014) 4470–4483. https://doi.org/10.1007/s11661-014-2370-6.

[72] M. Hillert, On the theory of normal and abnormal grain growth, Acta Metall. 13 (1965) 227–238. https://doi.org/10.1016/0001-6160(65)90200-2.

[73] F.J. Humphreys, M. Hatherly, The control of the grain size by particles, in: Recryst. Relat. Annealing Phenom., 2nd ed., Elsevier Ltd, 2004: p. 308.

[74] T. Kusama, T. Omori, T. Saito, S. Kise, T. Tanaka, Y. Araki, R. Kainuma, Ultra-large single crystals by abnormal grain growth, Nat. Commun. 8 (2017) 354. https://doi.org/10.1038/s41467-017-00383-0.

[75] P. Nandwana, M. Kirka, A. Okello, R. Dehoff, Electron beam melting of Inconel 718: effects of processing and post-processing, Mater. Sci. Technol. 34 (2018) 612–619. https://doi.org/10.1080/02670836.2018.1424379.

[76] D.P. Field, L.T. Bradford, M.M. Nowell, T.M. Lillo, The role of annealing twins during recrystallization of Cu, Acta Mater. 55 (2007) 4233–4241. https://doi.org/10.1016/j.actamat.2007.03.021.

[77] C.D. Barrett, A. Imandoust, A.L. Oppedal, K. Inal, M.A. Tschopp, H. El Kadiri, Effect of grain boundaries on texture formation during dynamic recrystallization of magnesium alloys, Acta Mater. 128 (2017) 270–283. https://doi.org/10.1016/j.actamat.2017.01.063.

[78] M. Zouari, N. Bozzolo, R.E. Loge, Mean field modelling of dynamic and post-dynamic recrystallization during hot deformation of Inconel 718 in the absence of δ phase particles, Mater. Sci. Eng. A 655 (2016) 408–424. https://doi.org/10.1016/j.msea.2015.12.102.

[79] A. Hadadzadeh, F. Mokdad, M.A. Wells, D.L. Chen, A new grain orientation spread approach to analyze the dynamic recrystallization behavior of a cast-homogenized Mg-Zn-Zr alloy using electron backscattered diffraction, Mater. Sci. Eng. A. 709 (2018) 285–289. https://doi.org/10.1016/j.msea.2017.10.062.

[80] I. Basu, T. Al-Samman, Twin recrystallization mechanisms in magnesium-rare earth alloys, Acta Mater. 96 (2015) 111–132. https://doi.org/10.1016/j.actamat.2015.05.044.

[81] A.A. Gazder, M. Sánchez-Araiza, J.J. Jonas, E.V. Pereloma, Evolution of recrystallization texture in a 0.78 wt.% Cr extra-low-carbon steel after warm and cold rolling, Acta Mater. 59 (2011) 4847–4865. https://doi.org/10.1016/j.actamat.2011.04.027.

[82] L.H. Rettberg, T.M. Pollock, Localized recrystallization during creep in nickel-based superalloys GTD444 and René N5, Acta Mater. 73 (2014) 287–297. https://doi.org/10.1016/j.actamat.2014.03.052.

[83] A. Imandoust, C.D. Barrett, A.L. Oppedal, W.R. Whittington, Y. Paudel, H.E. Kadiri, Nucleation and preferential growth mechanism of recrystallization texture in high purity binary magnesium-rare earth alloys, Acta Mater. 138 (2017) 27–41. https://doi.org/10.1016/j.actamat.2017.07.038.

[84] S. Mitsche, P. Pölt, C. Sommitsch, M. Walter, Quantification of the recrystallized fraction in a nickel-base-alloy from EBSD-data, Microsc. Microanal. 9 (2003) 344–345. https://doi.org/10.1017/S14319276030262229.

[85] S.W. Cheong, H. Weiland, Understanding a microstructure using GOS (grain orientation spread) and its application to recrystallization study of hot deformed Al-Cu-Mg alloys, Mater. Sci. Forum 558 (2007) 153–158. https://doi.org/10.4028/www.scientific.net/MSF.558-559.153.

[86] N. Allain-Bonasso, F. Wagner, S. Berbenni, D.P. Field, A study of the heterogeneity of plastic deformation in IF steel by EBSD, Mater. Sci. Eng. A 548 (2012) 56–63. https://doi.org/10.1016/j.msea.2012.03.068.

[87] A.K. Sinha, Growth of MC particles on stacking faults and dislocations, Metallography. 20 (1987) 37–45. https://doi.org/10.1016/0026-0800(87)90063-2.

[88] T. Omori, T. Kusama, S. Kawata, I. Ohnuma, Y. Sutou, Y. Araki, K. Ishida, R. Kainuma, Abnormal grain growth induced by cyclic heat treatment, Science 341 (2013) 1500–1502. https://doi.org/10.1126/science.1238017.

[89]  M.Y. Shen, X.J. Tian, D. Liu, H.B. Tang, X. Cheng, Microstructure and fracture behavior of TiC particles reinforced Inconel 625 composites prepared by laser additive manufacturing, J. Alloys Compd. 734 (2018) 188–195. https://doi.org/10.1016/j.jallcom.2017.10.280.
[90]  G.E. Dieter, D.J. Bacon, Dislocation sources, in: Mech. Metall., SI Metric, McGraw-Hill Book Company, 1986: p. 176.
[91]  M. Sundararaman, P. Mukhopadhyay, S. Banerjee, Deformation behavior of γ" strengthened Inconel 718, Acta Met. 36 (1988) 847–864. https://doi.org/10.1016/0001-6160(88)90139-3.
[92]  D. Fournier, A. Pineau, Low cycle fatigue behavior of Inconel 718 at 298 K and 823 K, Metall. Trans. A 8 (1977) 1095–1105. https://doi.org/10.1007/BF02667395.
[93]  S. Kondo, T. Mitsuma, N. Shibata, Y. Ikuhara, Direct observation of individual dislocation interaction processes with grain boundaries, Sci. Adv. 2 (2016) e1501926. https://doi.org/10.1126/sciadv.1501926.
[94]  K. Nishimoto, I. Woo, T. Ogita, M. Shirai, Mathematical simulation of grain boundary liquation in HAZ during welding-study on weldability of cast Alloy 718 rReport 5), Weld. Res. Abroad 19 (2001) 317–325. https://doi.org/10.2207/qjjws.19.317.
[95]  D.D. Pruthi, M.S. Anand, R.P. Agarwala, Diffusion of chromium in Inconel-600, J. Nucl. Mater. 64 (1977) 206–210. https://doi.org/10.1016/0022-3115(77)90026-5.
[96]  J. Čermák, Grain boundary self-diffusion of 51Cr and 59Fe in austenitic NiFeCr alloys, Mater. Sci. Eng. A. 148 (1991) 279–287. https://doi.org/10.1016/0921-5093(91)90830-G.
[97]  P. Lejcek, Grain boundary diffusion, in: Grain Bound. Segreg. Met., Springer, 2010: pp. 190–196. https://doi.org/10.1007/978-3-642-12505-8.

## Tables and Figures

Table 1. Nominal compositions of alloying elements in cast alloy and AM powders

| wt.% | Fe | Cr | Nb | Mo | Ti | Al | Mn | Co | Cu | Si | C |
|---|---|---|---|---|---|---|---|---|---|---|---|
| cast | 18.50 | 18.30 | 4.99 | 3.04 | 1.02 | 0.55 | 0.23 | 0.39 | 0.07 | 0.08 | 0.051 |
| AM | 18.26 | 18.87 | 4.97 | 2.97 | 0.94 | 0.46 | 0.06 | 0.23 | 0.05 | 0.06 | 0.03 |

Table 2. Sample notations and homogenization conditions of the present work

| Sample notations | Manufacturing methods | Homogenization conditions |
|---|---|---|
| AC | Suction-casting | - |
| AC20m | Suction-casting | 1180°C for 20 min with water quench |
| AC1h | Suction-casting | 1180°C for 1 h with water quench |
| AC12h | Suction-casting | 1180°C for 12 h with water quench |
| AB | AM-LPBF | - |
| AM20m | AM-LPBF | 1180°C for 20 min with water quench |
| AM1h | AM-LPBF | 1180°C for 1 h with water quench |
| AM12h | AM-LPBF | 1180°C for 12 h with water quench |

Table 3. Variation of the total phase fraction of both NbC and Laves_C14 phases during homogenization at 1180ºC.

| Homogenization time | 0 | 20 min | 1 hour | 12 hours |
|---|---|---|---|---|
| Suction casting | 13.13% ± 1.3% | 0.73% ± 0.07% | 0.65% ± 0.07% | 0.47% ± 0.05% |
| Additive manufacturing | 6.52% ± 0.7% | ~0 | 0.05% ± 0.005% | 0.36% ± 0.04% |

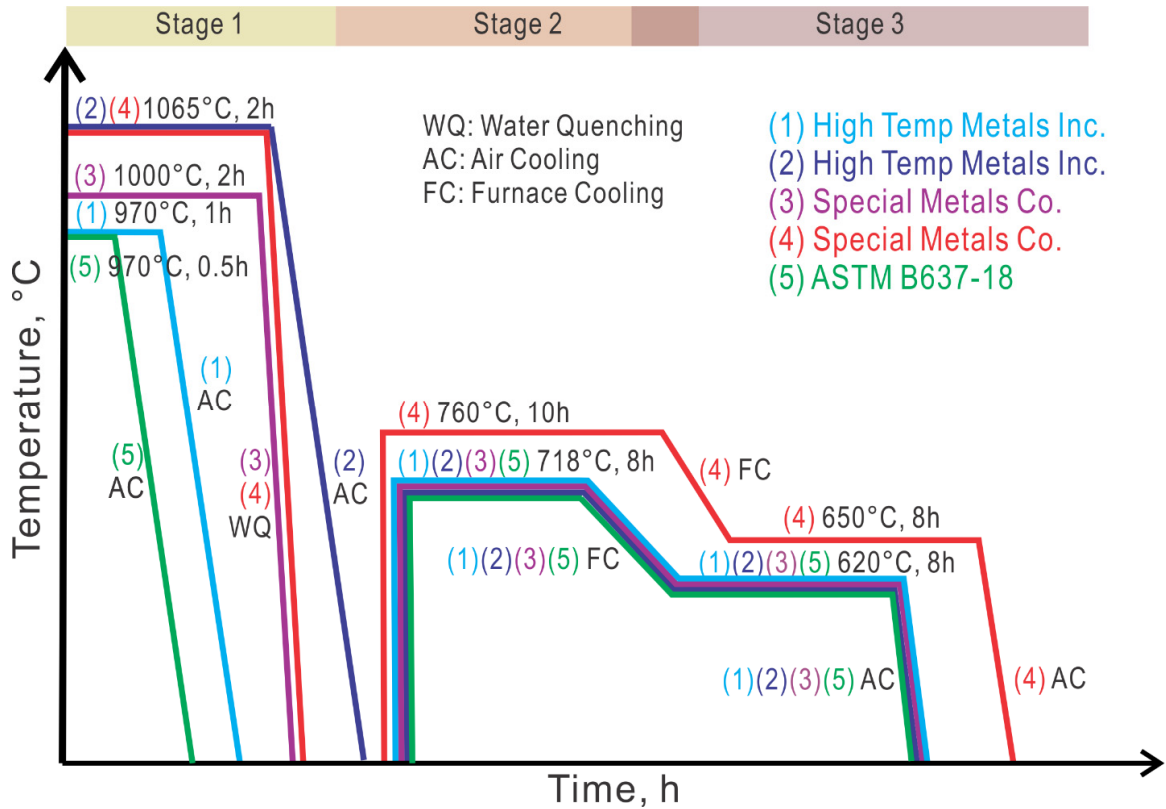

Figure 1. Available temperature profiles of heat treatment for the conventional Inconel 718 alloys.

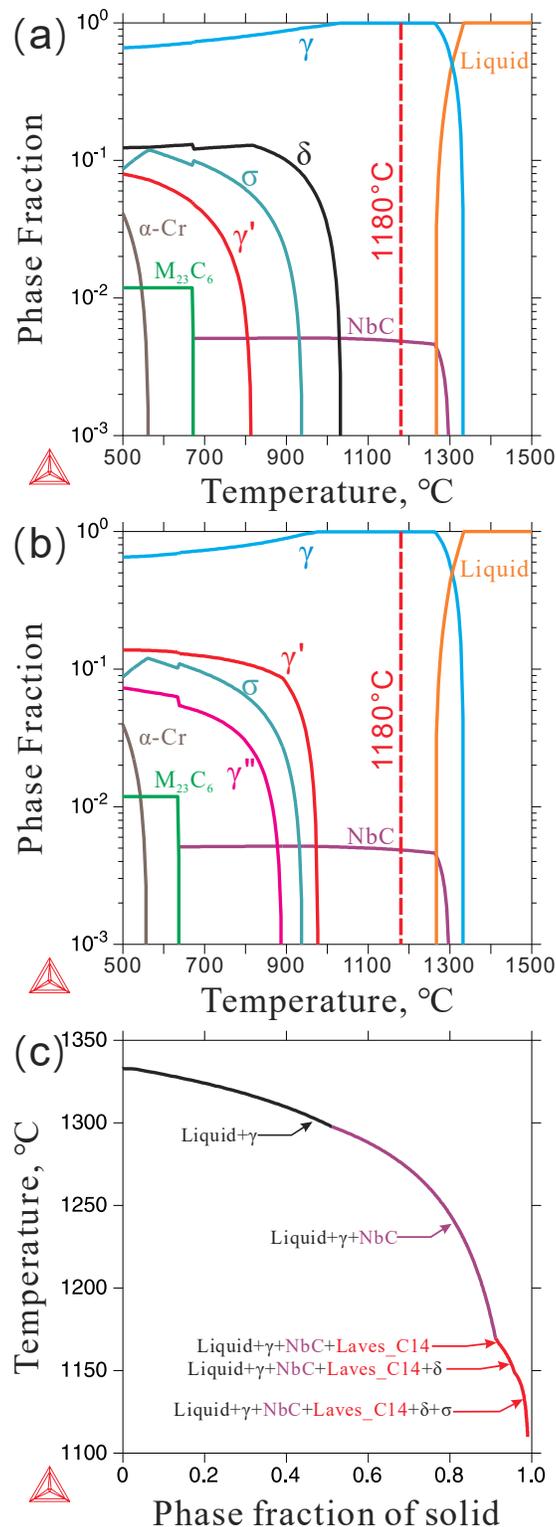

Figure 2. CALPHAD-based thermodynamic calculation for Inconel 718 based on the alloy composition Ni-18.5Fe-18.3Cr-4.99Nb-3.04Mo-1.02Ti-0.55Al-0.051C-0.23Mn-0.39Co-0.07Cu-0.08Si (in wt.%) of the suction cast alloy. (a) equilibrium step diagram; (b) nonequilibrium step diagram by suspending the δ phase; (c) nonequilibrium solidification paths predicted by the Scheil-Gulliver simulation.

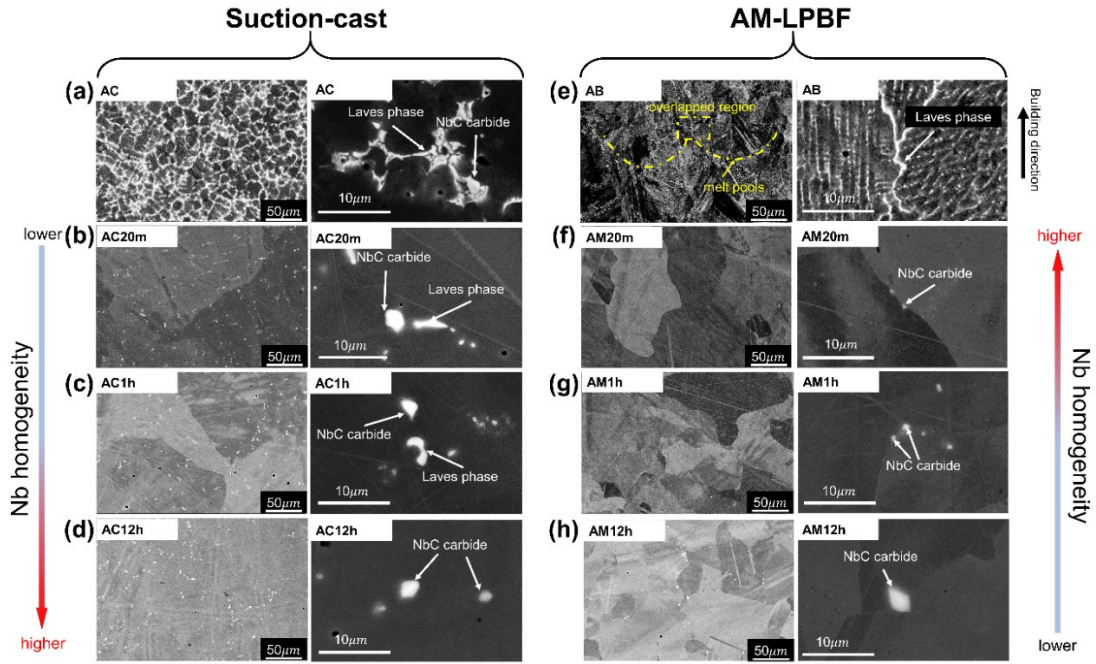

Figure 3. SEM-BSE micrographs of Inconel 718 suction-cast alloys with (a) as-cast state, AC; and with homogeneous states at 1180°C for (b) 20min, AC20m; (c) 1 h, AC1h; (d) 12 h, AC12h. SEM-BSE micrographs of AM-LPBF Inconel 718 alloys with (e) as-built state, AB; and with homogeneous states at 1180°C for (f) 20 min, AM20m; (g) 1 h, AM1h; (h) 12 h, AM12h. The representative NbC carbides and Laves_C14 phase in each sub-figure are shown in the magnified SEM images on the right-hand side thereof.

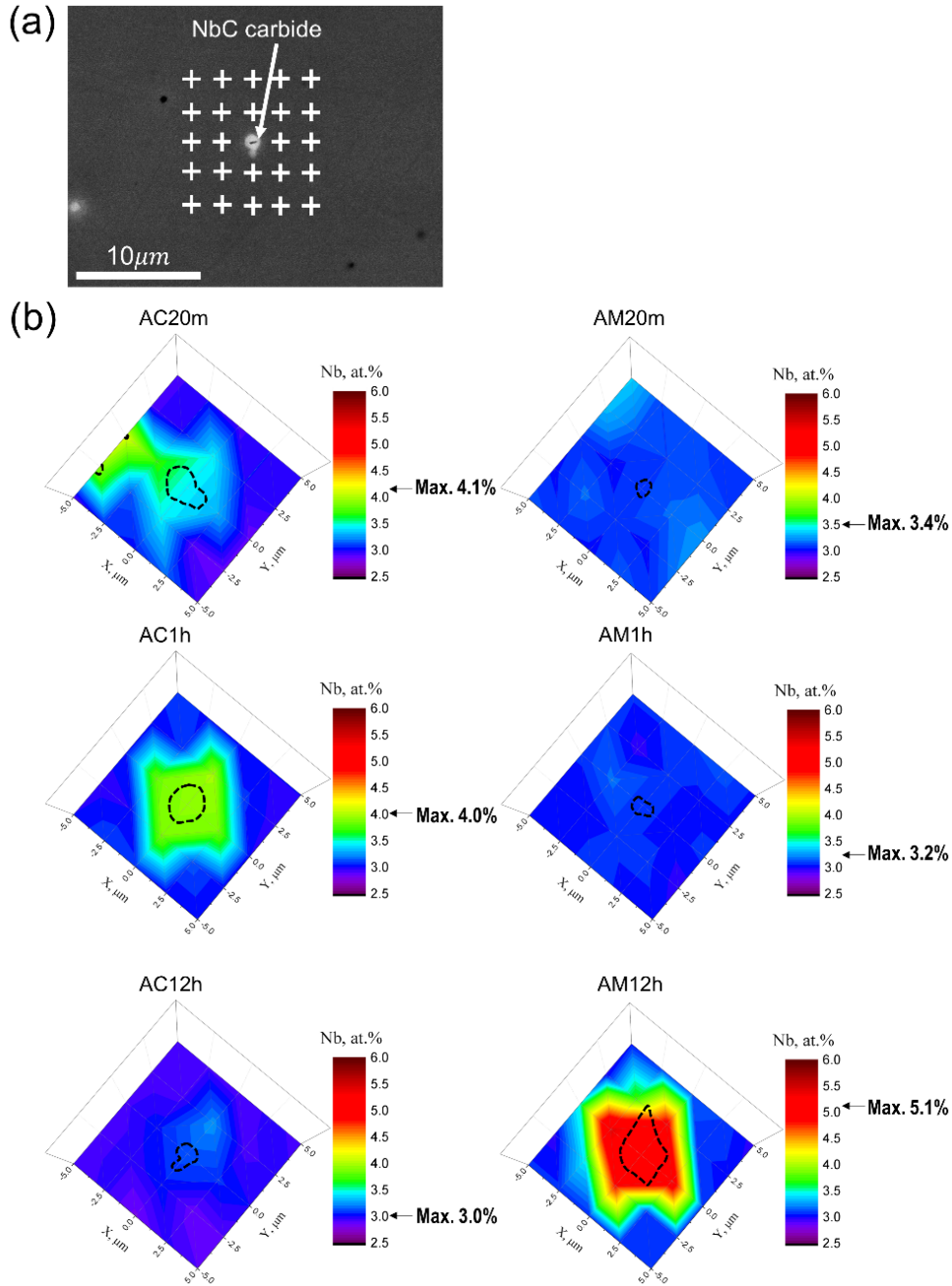

Figure 4. (a) Illustration of EDS point identification for Nb homogeneity determination (taking sample AC12h, homogenized suction-cast alloy at 1180°C for 12 h as an example); (b) Nb concentration contour diagrams (10×10 μm) in γ matrix around NbC carbides for presenting Nb homogeneity. The NbC carbides are profiled by dashed black squiggles.

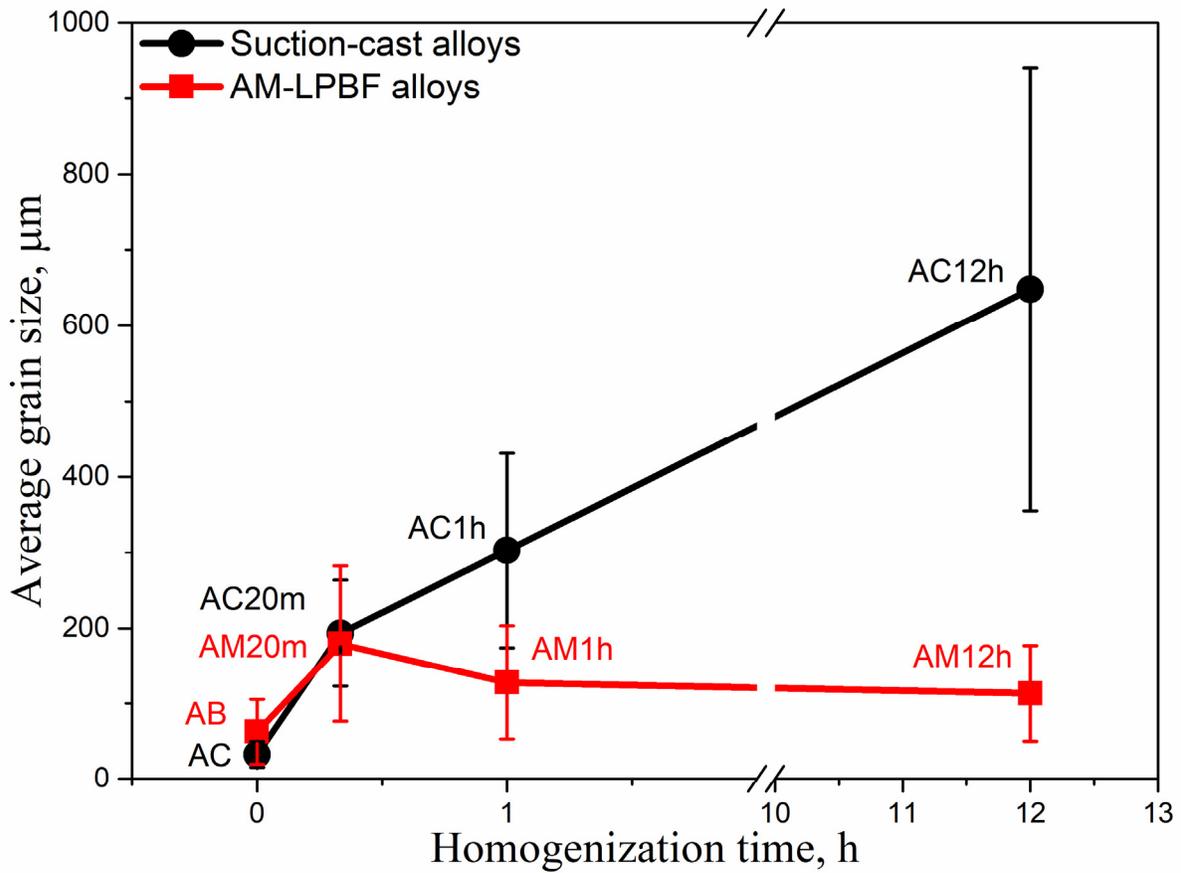

Figure 5. The average grain size of suction-cast and AM-LPBF alloys before and after homogenization at 1180°C.

Yunhao Zhao, Kun Li, Matthew Gargani, Wei Xiong, *Additive Manufacturing*, 36 (2020) 101404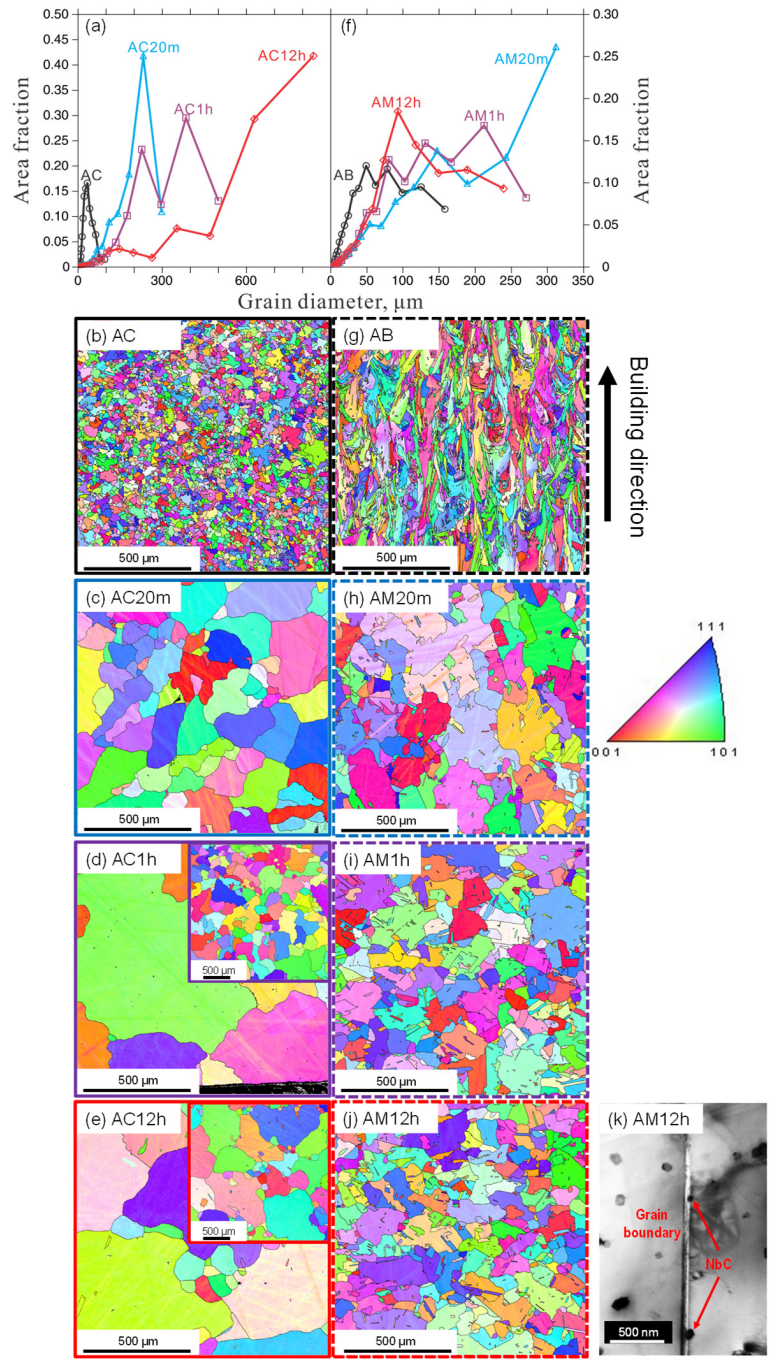

Figure 6. Grain size distribution of (a) suction-cast alloys; and inverse pole figure (IPF) orientation maps from EBSD of suction-cast Inconel 718 alloys with (b) as-cast state, AC; and with homogeneous states at 1180°C for (c) 20 min, AC20m; (d) 1 h, AC1h; (e) 12 h, AC12h. Grain size distribution of (f) AM-LPBF alloys; and IPF orientation maps of AM-LPBF Inconel 718 alloys with (g) as-built state, AB; and with homogeneous states at 1180°C for (h) 20 min, AM20m; (i) 1 h, AM1h; (j) 12 h, AM12h. (k) TEM micrograph of sample AM12h.

Page **26** of **28**

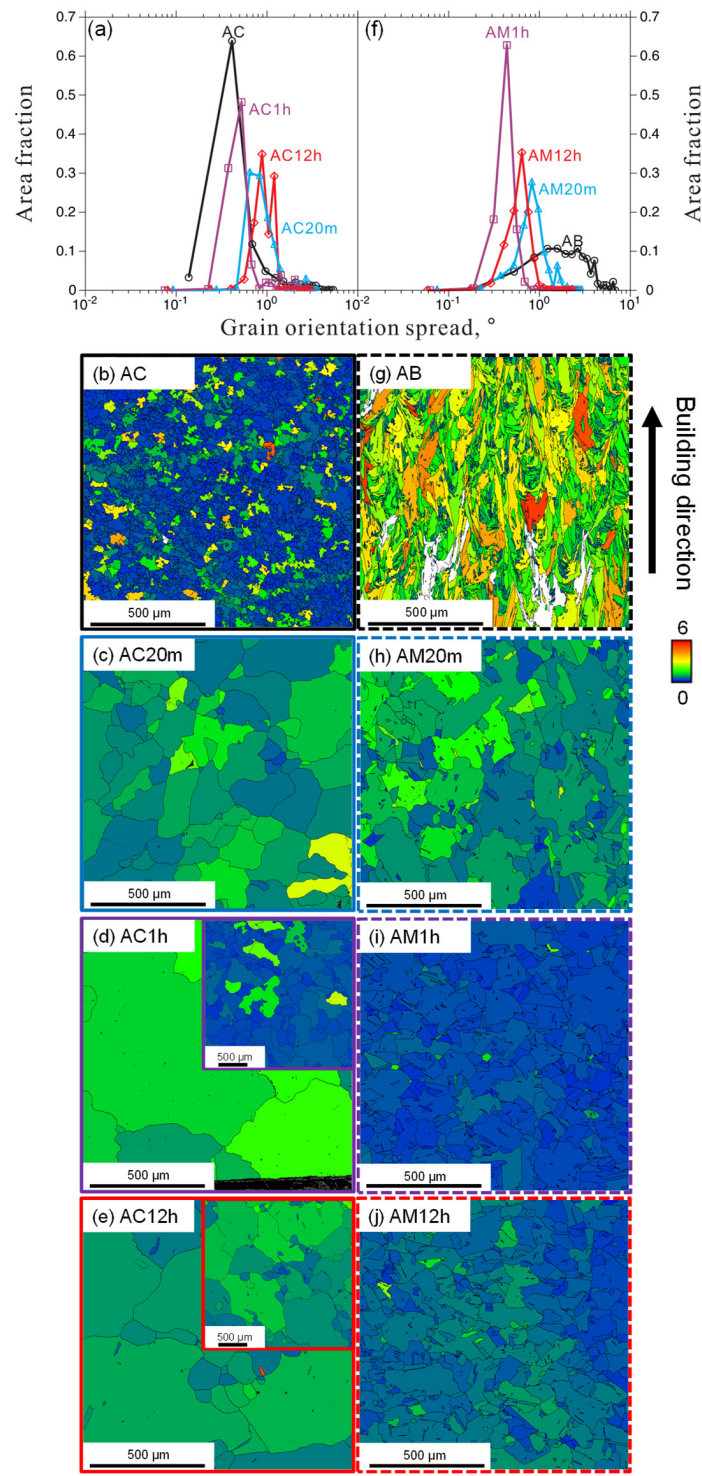

Figure 7. GOS distribution of (a) suction-cast alloys; and grain orientation spread (GOS) maps from EBSD of suction-cast Inconel 718 alloys with (b) as-cast state, AC; and with homogeneous states at 1180°C for (c) 20 min, AC20m; (d) 1 h, AC1h; (e) 12 h, AC12h. GOS distribution of (f) AM-LPBF alloys; and GOS maps of AM-LPBF Inconel 718 alloys with (g) as-built state, AB; and with homogeneous states at 1180°C for (h) 20 min, AM20m; (i) 1 h, AM1h; (j) 12 h, AM12h.

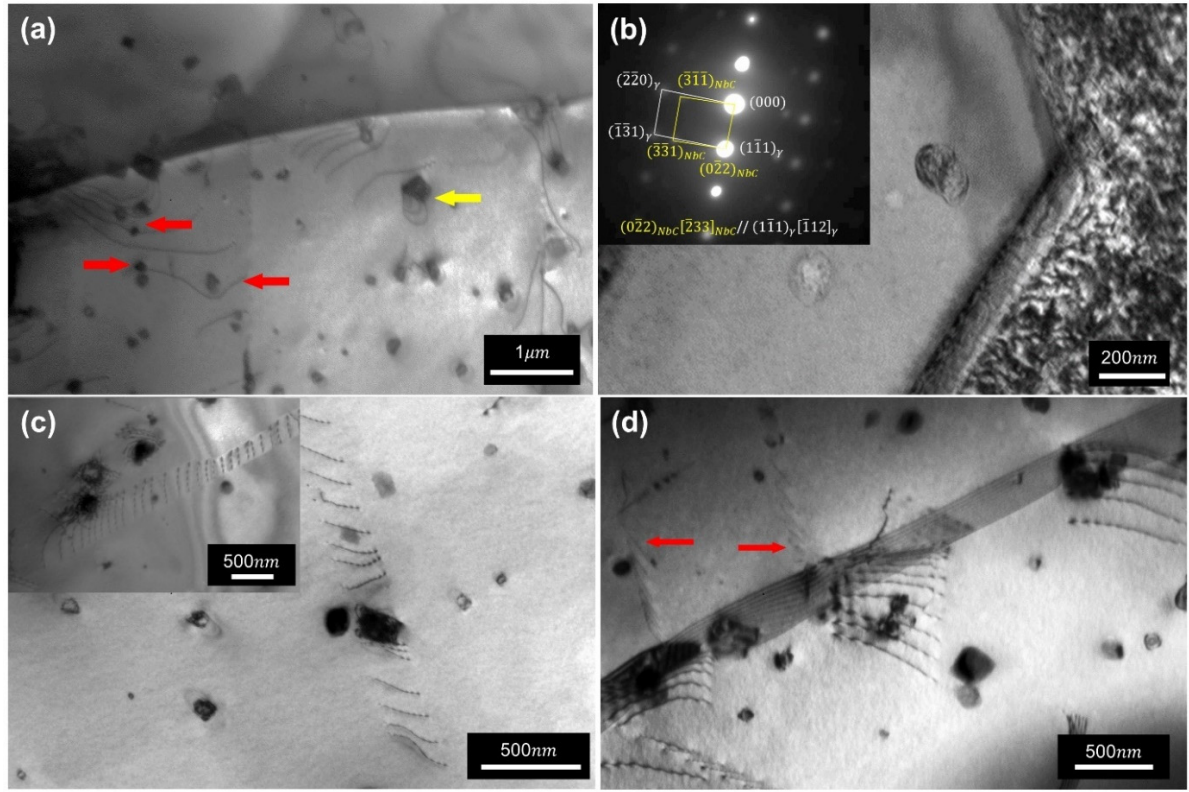

Figure 8. TEM micrographs of sample AM12h (homogenized on AM-LPBF alloy at 1180°C for 12 h): (a) grain boundary NbC carbides and dislocation-precipitate interaction; (b) selected area electron diffraction pattern of NbC carbides; (c) planar dislocation arrays; (d) interaction between dislocations and grain boundaries.